\documentclass[a4paper,12pt,fleqn,titlepage,onecolumn,oneside,final,bibliography=totocnumbered,numbers=noenddot]{article}
\usepackage{graphicx}
\usepackage{subcaption}
\usepackage{natbib}
\usepackage[utf8]{inputenc}
\usepackage[english]{babel}
\usepackage{url}
\usepackage{caption}
\usepackage{booktabs}
\usepackage[pagewise]{lineno}

\usepackage[a4paper,left=1.0in,right=1.0in,top=0.5in,bottom=0.5in,includeheadfoot]{geometry}

\title{Investigating the potential of social network data for transport demand models}
\date{June 30, 2017}
\author
{Michael A.B. van Eggermond,$^{1\ast}$ Haohui Chen,$^{2,3}$
	\and 
	Alexander Erath,$^{1,4}$ Manuel Cebrian$^{2}$\\
	\\
	\normalsize{$^{1}$Future Cities Laboratory, Singapore ETH Centre, Singapore}\\
	\normalsize{$^{2}$Data61, Commonwealth Scientific and Industrial Research Organization, Australia}\\
	\normalsize{$^{3}$Monash University, Australia}\\
	\normalsize{$^{4}$IVT, ETH Zürich, Switzerland}\\
	\\
	\normalsize{$^\ast$To whom correspondence should be addressed; E-mail:  eggermond@ivt.baug.ethz.ch.}
}

\begin{document}
\maketitle

\section*{Abstract}
Location-based social network data offers the promise of collecting the data from a large base of users over a longer span of time at negligible cost. While several studies have applied social network data to activity and mobility analysis, a comparison with travel diaries and general 
statistics has been lacking.

In this paper, we analyzed geo-referenced Twitter activities from a large number of users in Singapore and neighboring countries. By combining this data, population statistics and travel diaries and applying clustering techniques, we addressed detection of activity locations, as well as spatial separation and transitions between these locations. 

We had large numbers of Twitter users in the data set collected over a period of 8 months; however, due to the scattered nature of the data, only a group comparable to a travel survey turned out to be useful for further analysis. Kernel density estimation performs best to detect activity locations; more activity locations are detected per user than reported in the travel survey.

The descriptive analysis shows that determining home locations is more difficult than detecting work locations for most planning zones. Spatial separations between detected activity locations from Twitter data - as reported in a travel survey and captured by public transport smart card data - are mostly similarly distributed, but also show relevant differences for very short and very long distances. This also holds for the transitions between zones.

Whether the differences between Twitter data and other data sources stem from differences in the population sub-sample, clustering methodology, or whether social networks are being used significantly more at specific locations must be determined by further research.

Despite these shortcomings, location-based social network data offers a promising data source for insights in activity locations and mobility patterns, especially for regions where travel survey data is not readily available.

\newpage



%
\section{Introduction}
The well-established four-step transport model, as well as state-of-the art agent-based models, have relied on the same data sources over the last several decades. Traditional data sources included, but were not limited to, travel diary surveys, population censuses, business censuses, road networks and transit schedules.

Travel diary surveys collect information on individuals and their households; along with trip information, they also cover information on activities and locations visited over the course of one day. Population and business censuses provide insight into (aggregate) population statistics by home and work location. These and other strands of data are then combined to quantify travel demand for a general population used in transport models. 

Shortcomings of travel diaries include the common under-reporting of short trips; it is also difficult to sample from all potential user groups over a longer time span in study regions due to time and budget limitations. Furthermore, both travel diary surveys and censuses were usually conducted every 5 to 10 years, with access for researchers and the general public sometimes restricted.


Social network data offers the chance to observe users over a larger time span for very low costs. Previous research investigated activity and mobility patterns \citep{HasanSEtAl_UrbComp_2013} a) and focused on the recognition of mobility patterns in a range of cities  \citep{NoulasEtAl_PLOS_2012}. These studies have shown the possibilities of using social network data; however, a comparison with travel diaries or other transport related data sources was lacking. 

In this paper, we investigate the possibilities of using publicly available social network data for transport planning purposes. More specifically, we try to investigate the possibility of using Twitter data’s spatial and temporal information  to complement, or replace, travel diaries. We collected data from the social networking and micro-blogging service Twitter for 8 months for Singapore and neighboring countries Malaysia, Indonesia and Thailand. To test the validity of travel patterns derived from this data, we compared it to Singapore’s household interview travel survey and one week of public transport smart card data. By merging these data sources and applying several clustering methods, we addressed the following questions:

\begin{enumerate}
  \item Is it possible to recognize activity locations from social network data and, if so, do the detected activity locations correspond to activity locations reported in other, more traditional transport data sources? 
  \item Is the spatial separation between detected activity locations comparable to distances between reported activity locations from travel diaries? 
  \item Is it possible to derive origin destination matrices from social network data and how do these matrices correspond to observed trips? 
\end{enumerate}

The remainder of this paper is structured as follows. Section \ref{sec:lit_review} continues with a literature review. Section \ref{sec:data} provides an overview of the used data sources, methodology is presented in section \ref{sec:methodology}and results are presented in section \ref{sec:results}. Section \ref{sec:discussion} concludes with the discussion and outlook. 
\section{Background}\label{sec:lit_review}
\subsection{Data collection for transport planning and modeling}
The normal method for obtaining travel behavior data was a one-day travel diary survey, which collected data for a representative day with the representative peak hour \cite[e.g.][]{OrtuzarWillumsen_2011}. In travel diaries, data is gathered about household, person and travel behavior characteristics, including household income, residential location, dwelling type, etc. Trip data includes a start time, end time, mode(s) used, number of transfers and trip purpose. Activity duration is derived from the difference between the trip end time and trip start time of subsequent trips.

In the case of the classic four-step model \cite[e.g.][]{OrtuzarWillumsen_2011} and, more recently, activity-based models \cite[e.g.][]{AxhausenEtGaerling_TransportRev_1992}, this data was used for trip generation, activity chain generation, trip distribution, modal choice and traffic assignment models. In addition to these models, information was required about zonal productions and zonal attractions (e.g. number of household, workplaces, leisure locations). 

While travel diary data often effectively captured travel demand patterns, it is well known that data collection method details (like exact departure time and arrival, route taken and number of activity locations) hindered adequate surveys. To obtain more detailed user data, state-of-the-art travel surveys included, or were supplemented by, GPS data.

Subsequently, numerous studies have assessed the accuracy of one-day travel surveys. \cite{Wolf_ISCTSC_2004} reported that the rate of missing trips ranged from 11\% to 81\%, when comparing the results of computer assisted telephone surveys and GPS data in six household surveys in the US. \cite{ForrestPearson_TRR_2005} found that the number of trips observed in GPS data in their 2002 survey was much higher than the number of trips reported by respondents. Home-based non-work and non-home-based trips suffered most from under-reporting. In both these studies, in-vehicle 
GPS units were used to assess the accuracy of survey response. \citet{StopherEtAl_Transportation_2007} found a lower  percentage of missing trips when comparing results of a travel survey conducted by face-to-face interviews and hand-held GPS devices. In their analysis of missing trips, they found that people performing many trips under-report them. Other problems occurred with trips conducted after 5pm, non-motorized trips and activities with a duration under 10 minutes. 

Transport planners and researchers know that one-day travel surveys do not capture the full range of activities by individuals and households, such as social visits, leisure activities and shopping. 

Since the Uppsala Travel Survey of 1971, a number of multi-day and multi-week travel surveys have been conducted, despite the higher costs, survey fatigue and low response rates typical for longer surveys. \cite{SchlichAxhausen_Transportation_2003} introduced different methods to measure the similarity of day-to-day travel behavior in a six-week travel survey conducted in Karsruhe and Halle, Germany \citep{AxhausenEtAl_Transportation_2002}. They showed that behavior is neither  totally repetitious, nor totally variable. The different methods employed to measure similarity showed that travel is more stable on workdays and that at least two weeks of data was required to measure similarity.
 
An increasing number of cities, regions and countries have adopted public transport smart cards. While the main objective of these systems is to collect revenue, another result is very detailed data of onboard transactions that can be used for numerous applications \citep{PelletierEtAl_TransResC_2011}. Depending on the type of implementation of the smart card system, a trip start time and/or end time are available to the transport company. Disadvantages of smart card data include the lack of trip purpose and the lack of knowledge about exact origin and destination location of a public transport user’s trip \citep{BagchiWhiteP_TPol_2005}. Despite these disadvantages, it is still possible to extract trip distance, trip duration (excluding waiting time) and an individual’s approximate time spent at a location \citep[e.g.][]{ChakirovErath_IATBR_2012}. 

Several studies have already explored the possibility of using mobile phone Call Detail Records (CDR) to estimate origin destination matrices \cite{CalabreseEtAl_PC_2011,WangEtAl_ICICTA_2011,IqbalEtAl_TransResC_2014}. \citet{IqbalEtAl_TransResC_2014} point out that, while it is possible to extract trip patterns from mobile phone data, heterogeneity exists in call rates from different locations, leading to biased results.

\subsection{Social network data}
Social network services build on the real-life social networks of people through online platforms to share ideas, activities and interests; the increasing availability of location-acquisition technology offers the extra possibility for people to add a location dimension to existing social networks in various ways \citep{Zheng_2011}.

Within the field of transport modeling, location-based social network data has been used to classify users' activity patterns \citep{HasanSUkkusuri_TransResC_2014}, detected traffic anomalies \citep{PanEtAl_SIGSPATIAL_2013}, reconstructed popular traffic routes \citep{WeiEtAl_KDD_2012}, recognized mobility patterns in a range of cities \citep{NoulasEtAl_PLOS_2012} and modeled human location  \citep{LichmanSmyth_KDD_2014}. 

While an increasing number of studies use geo-tagged social network data, less attention is being paid to the social network data’s representation of the general population \citep{Grossenbacher_MastersThesis_2014}. One analysis stated \citep{Haklay_2012}:

\emph{In digiplace the wealthy, powerful, educated and mostly male elite is amplified through multiple digital representations. Moreover, the frequent decision of algorithm designers to highlight and emphasise those who submit more media, and the level of 'digital cacophony' that more active contributors create, means that a very small minority - arguably outliers in every analysis of normal distribution of human activities - are super empowered.}

However, location-based social network data comes with a larger sample size for a longer period without any significant costs \citep{HasanSUkkusuri_TransResC_2014}. Several disadvantages limit the use of traditional econometric tools for these data sets \citep{HasanSUkkusuri_TransResC_2014}: (i) they do not provide detailed descriptions of activities, such as start and end times and activities can be either at static locations or en-route (ii) individuals are recognized only by an identifier without additional information on individual socio-economic characteristics; (iii) the data has missing activities, since the only activities observed are those an individual shares in social media. In addition to this last point, it should also be noted that only users active in social media are included.
\section{Data collection \& preparation}\label{sec:data}
\subsection{Study region}
The availability of a travel diary survey, public transport smart card data and a broad Twitter user base made Singapore an ideal case study to answer our research questions. Singapore is located in Southeast Asia with a land area of 712 km$^2$, a permanent residential population of 3.77 million and a total population of 5.08 million in 2010, compared to respectively 697 km$^2$, 
3.27 million and 4.03 million in 2000. GNI per capita amounts to US\$ 54,580, 2013), which makes it one of the wealthiest countries in (Southeast) Asia. 

\subsection{Social network data}
The social networking and microblogging service Twitter was launched in 2006. At the end of March 2014, there were 255 million average monthly active users (MAUs), of which 198 million  mobile MAUs \citep{Twitter_FQ_2014}. Together, these users send 500 million tweets per day \citep{Twitter_Webpage_2014}. 

As opposed to many other social networking sites, Twitter offers the opportunity to download users’ profiles and Twitter messages, or tweets, including the geo-location of the tweet; it also includes an indicator whether it was sent from a mobile device, or from a computer in real-time. 

When downloading data from Twitter, one can specify a geographic area. For this research, we have specified the bounding box 'Singapore', shown in Figure 1. In total, 4,121,433 tweets were collected during the period from September 10, 2013 to February 27, 2014. While a geographic bounding box has been specified, not 
all tweets are geo-tagged with a longitude and latitude and not all tweets are located in Singapore. Table \ref{tab:general_overview}  lists the number of users, tweets, geo-tagged tweets and tweets in Singapore. 

Additionally, an indicator has been included to see whether a user has tweeted 10 times or more within the time span specified earlier. One sees that only 29\% of users Tweet 10 times or more within this period, but that these users contribute over 90\% of the tweets. Table \ref{tab:general_overview}  highlights these statistics and provides a range of other figures on smart card data, the Singaporean travel diary survey and Twitter data. These figures will be discussed later in this section. 
 
\begin{table}[htbp]
	\scriptsize
  \centering
  \caption{Aggregates from different data sources}
  
    \begin{tabular}{lrrr}
\toprule
    \textbf{Data source / Indicator} & \multicolumn{1}{l}{\textbf{All users}} & \multicolumn{1}{l}{\textbf{10 tweets or more}} & \multicolumn{1}{l}{\textbf{Percentage}} \\
    \midrule
    \multicolumn{1}{l}{\textit{Twitter (from September 10, 2013 until February 27, 2014)}} &       & \multicolumn{1}{l}{\textit{}} & \multicolumn{1}{l}{\textit{}} \\
    \multicolumn{1}{l}{Number of users} &                    157,043  &                  45,715  & 29.1 \\
    \multicolumn{1}{l}{Number of tweets} &                  4,121,433  &              3,800,904  & 92.2 \\
    \multicolumn{1}{l}{Number of geo-tagged tweets} &                  3,703,425  &              3,417,418  & 92.3 \\
    \multicolumn{1}{l}{Number of tweets in Singapore} &                  2,129,930  &              1,957,952  & 91.9 \\
    \multicolumn{1}{l}{Number of tweets outside Singapore} &                  1,573,495  &              1,459,466  & 92.8 \\
    \multicolumn{1}{l}{Number of users tweeting only in Singapore} &                      77,234  &                  20,822  & 27.0 \\
    \multicolumn{1}{l}{Number of users tweeting only outside Singapore} &                      54,682  &                  14,528  & 26.6 \\
    \multicolumn{1}{l}{Number of users tweeting in Singapore and overseas} &                        9,189  &                    5,857  & 63.7 \\
    \multicolumn{1}{l}{} &       &       &  \\
    \multicolumn{1}{l}{\textit{Household interview travel survey 2012}} &       &       &  \\
    \multicolumn{1}{l}{Number of households} &                      9,635  &       &  \\
    \multicolumn{1}{l}{Number of persons} &                      35,714  &       &  \\
    \multicolumn{1}{l}{} &       &       &  \\
    \multicolumn{1}{l}{\textit{Smart card data (from April 6 to April 12, 2014)}} &       &       &  \\
    \multicolumn{1}{l}{Number of card identifiers in smart card data} &                  3,475,574  &       &  \\
    \multicolumn{1}{l}{Number of journeys over 7 days} &                23,994,771  &       &  \\
    \multicolumn{1}{l}{} & \multicolumn{1}{l}{} & \multicolumn{1}{l}{} & \multicolumn{1}{l}{} \\
    \multicolumn{1}{l}{\textit{Singapore statistics (2012 except where otherwise stated)}} & \multicolumn{1}{l}{} & \multicolumn{1}{l}{} & \multicolumn{1}{l}{} \\
    \multicolumn{1}{l}{Total population} & 5,312,400 & \multicolumn{1}{l}{} & \multicolumn{1}{l}{} \\
    \multicolumn{1}{l}{Total resident population (Singaporeans and Permanent Residents)} & 3,818,200 & \multicolumn{1}{l}{} & \multicolumn{1}{l}{} \\
    \multicolumn{1}{l}{Singaporeans} & 3,285,100 & \multicolumn{1}{l}{} & \multicolumn{1}{l}{} \\
    \multicolumn{1}{l}{Permanent resident} & 533,100 & \multicolumn{1}{l}{} & \multicolumn{1}{l}{} \\
    \multicolumn{1}{l}{Total non-resident population (Workpass holders \& dependents)}  & 1,494,200 & \multicolumn{1}{l}{} & \multicolumn{1}{l}{} \\
    \multicolumn{1}{l}{Land-area 2013 [km2]} & 716.1 & \multicolumn{1}{l}{} & \multicolumn{1}{l}{} \\
    \multicolumn{1}{l}{Population density 2013 [persons per km2] (Based on total population)} & 7,540 & \multicolumn{1}{l}{} & \multicolumn{1}{l}{} \\
    \multicolumn{1}{l}{Per capita GNI 2013 [US\$] }& 54,580 &       &  \\
    \bottomrule

    \end{tabular}%
  \label{tab:general_overview}%
\end{table}%

\begin{figure}
  
  \centering
\includegraphics[width=0.9\textwidth]{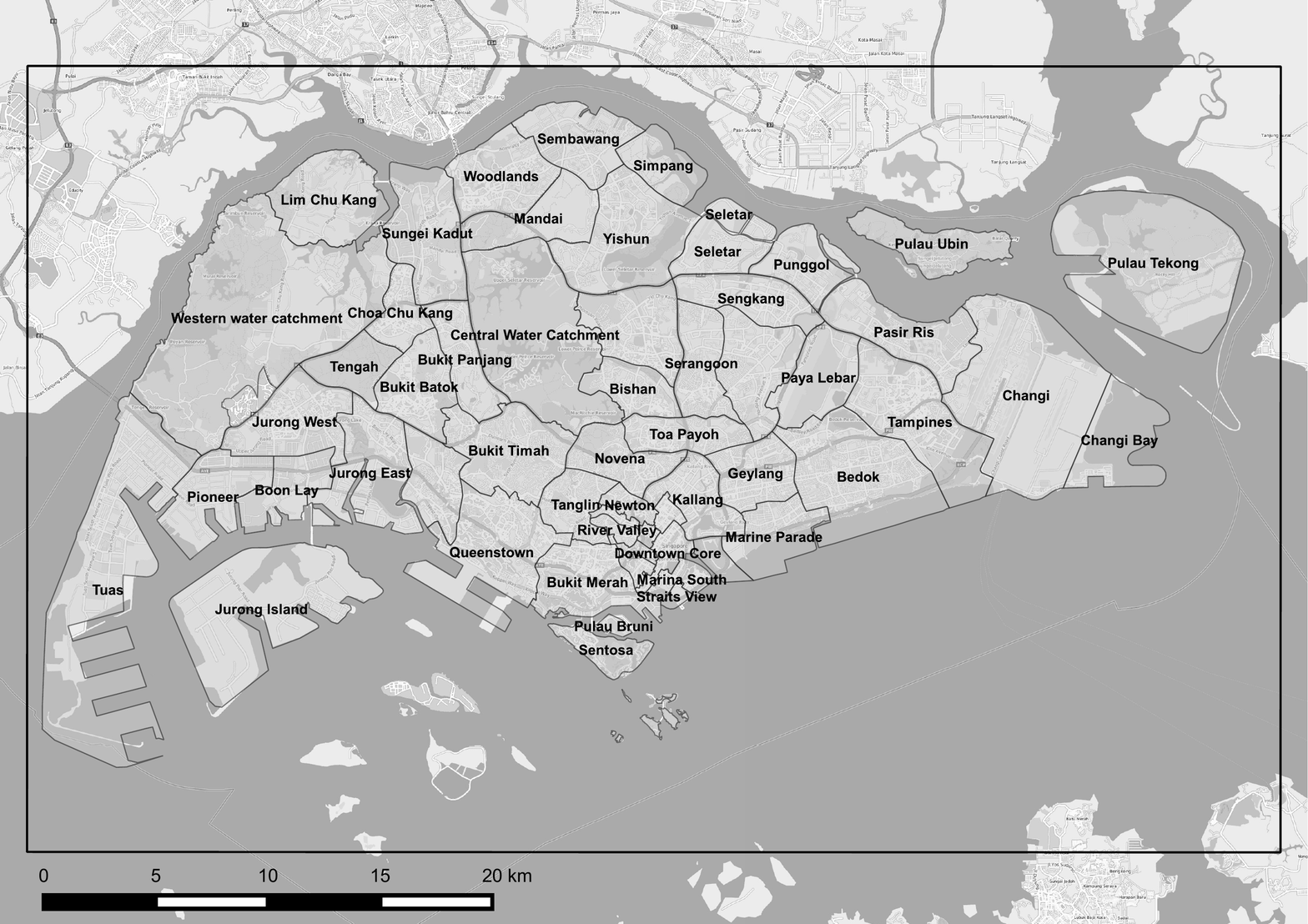}
\caption{Specified bounding box and planning zones. Background map: \protect\url{http://openstreetmap.org}.}
  \label{fig:planning_zones}
\end{figure} 
\subsection{Smart card data}
Singapore’s public transport card was introduced in April 2002; smart cards can be used island-wide for payment of all modes of public transport, regardless of operator. Though cash payment of single fares at higher rates is still possible, e-payments using smart cards account for 96\% of all trips \citep{Prakasam_JRTR_2008}. Smart card data records include the boarding station, boarding time, end station and time. In this paper, we employ 7 days of smart card data from trips made between April 6 to April 12, 2014. 
\subsection{Travel diary survey}
Trip information is taken from the Household Interview Travel Survey (HITS) 2012. For this survey, over 32,000 persons from over 9,000 households reported their travel behavior on a single workday, corresponding to approximately 1\% of households in Singapore. The survey is conducted once every four years and is commissioned by the Singaporean Land Transport Authority (LTA). HITS contains data on three levels of aggregation; the highest level contains household characteristics. In the second highest level, personal characteristics are available, such as age, income, profession and employment type. On the lowest level of aggregation, trip information is available for mode, purpose, cost and time. 
\subsection{Other data sources}
We enriched the previous data sets with attributes from several other data sources, including the 1,092 transport analysis zones (TAZ’s) as used in Singapore’s transport demand model, the 55 land-use planning zones and land-use types. Singapore’s populations statistics \citep{SingStat_Webpage_2010} were included, as well as an estimated number of work activities in Singapore by planning zone \citep{ChakirovErath_IATBR_2012, OrdonezErath_TRR_2013}.

\section{Methodology}\label{sec:methodology}
\subsection{Identification of clusters}
To assess the suitability of Twitter data for transport demand analysis, we wanted to recognize stationary activity locations visited by an individual: his or her home location, work location, education locations and locations where discretionary activities were performed. As such, we did not touch on the fact that activities can also be performed en-route. For instance, 	it is possible to work while commuting or maintain social contacts. By observing an individual over a longer span of time it would be possible to capture more activity locations than from a one or two day household travel survey. To differentiate between stationary activities and activities performed en-route, we assumed that events (tweets) posted at activity locations tended to be less geographically dispersed; en-route events would be more geographically dispersed.
Partitioning geographically close activities into clusters helped identify en-route activities, as their clusters should contain fewer events. 


\subsection{K-means clustering}
K-means clustering is one of the most popular clustering methods \citep{JainA_PRL_2010}. Since K-means clustering was proposed in 1955, many studies have applied different variations of the method in a wide range of domains. Finding the optimal number of clusters $k$ is a challenging, but necessary task. Various ways of obtaining the optimal $k$ value were discussed in \citet{JainA_PRL_2010}, essentially trying different values of $k$ and selecting the best value based on predefined criteria, such as the minimum message length \citep{FigueiredoJainA_IEEETPAMI_2002}, minimum description length \citep{HansenMHYuB_JASA_2001}, gap statistics \citep{TibshiraniEtAl_JORSSB_2001}  and Dirichlet process \citep{RasmussenC_SollaEtAl_2000}. 

A more general and easy-to-implement method for validating clustering results was the silhouette method \citep{Rousseeuw_JACM_1987}. The value of a silhouette measured (1) how well an observation was assigned to its cluster and (2) how dissimilar that observation was to its neighboring clusters, thus reflecting the clustering analysis performance. This paper used the value of the silhouette to validate the clustering results of different $k$ values and selected the optimal value. 

\begin{figure}
  
  \centering
\includegraphics[width=0.5\textwidth]{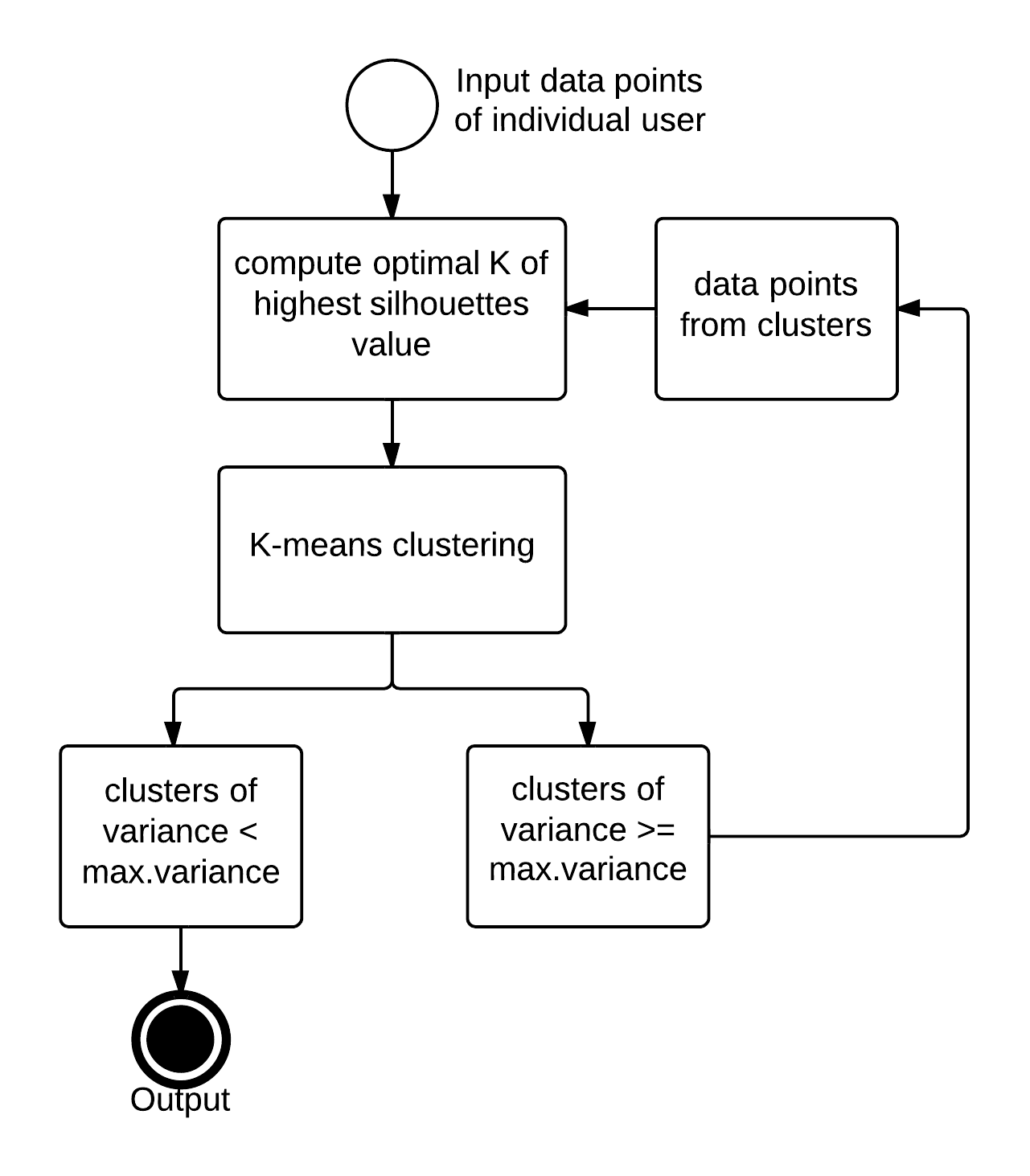}
\caption{K-means clustering procedure}
  \label{fig:clustering_procedure}
\end{figure}

Clusters resulting from k-means clustering can be fairly large if measured by the convex hull of all the events (tweets) included in the cluster. For this research, we asserted that a large cluster cannot necessarily constitute a single activity location. We defined a maximum threshold for the variance of 200 meters. Clusters which exceeded this threshold were recursively broken down into more, smaller clusters by recursive k-means clustering \citep{AshbrookStarner_PUC_2003}. This process is also illustrated in Figure \ref{fig:clustering_procedure}. 

\subsection{Kernel density estimation and clustering}
Kernel density estimation (KDE) provided us with another way of determining activity locations that individuals frequently visited. It is a non-parametric method for estimating a density function from a random sample of data \citep{SilvermanB_2001}. A user-defined parameter called bandwidth $h$ specifies the standard deviation of the Gaussian distribution function constructed around each data point to smooth the KDE result. A small value for $h$ may under-smooth the KDE result; a large value for $h$ can result in over-smoothing. The bandwidth $h$ gives an indication of the area covered by the KDE result; 95\% will be located within 400 meters.

Two main methods for the selection of bandwidth exist: the fixed bandwidth and the adaptive bandwidth method. Given that each location should be limited in area, a universal fixed bandwidth for the KDE is assumed, which is the same as the k-means clustering's maximum variance, namely $h$ = 200 meters. To obtain clusters from the estimation procedure, contour lines are constructed. All local peaks of the contour line are regarded as clusters and contour levels are assigned to corresponding kernels. The resulting levels are calculated per cluster per individual and provide an indication of the relative importance of a single cluster among an individual's clusters; cluster levels between different individuals cannot be directly compared. 

KDE, by itself, is not a clustering method. However, as the clusters (peaks) are impacted by neighboring activities, neighboring activities within a certain distance ($h$) are grouped together. If an activity (tweet) belongs to more than one kernel, it is grouped to the closest one. This situation is very rare in our data set and occurs in less than 0.01\% of cases. 
\subsection{Limitations of applied clustering methods}
Not all recorded tweets can be assigned to a cluster. Individual tweets sent from locations that are visited only rarely, for instance during a restaurant visits, or at a concert, would typically not be assigned to a cluster. Therefore, our approach did not assign these tweets to activity locations, just discarded them. Furthermore, the applied approach did not make use of a tweet's temporal attribute; using this component provided an additional challenge. 

\begin{figure}
 
  \centering
\includegraphics[width=0.8\textwidth]{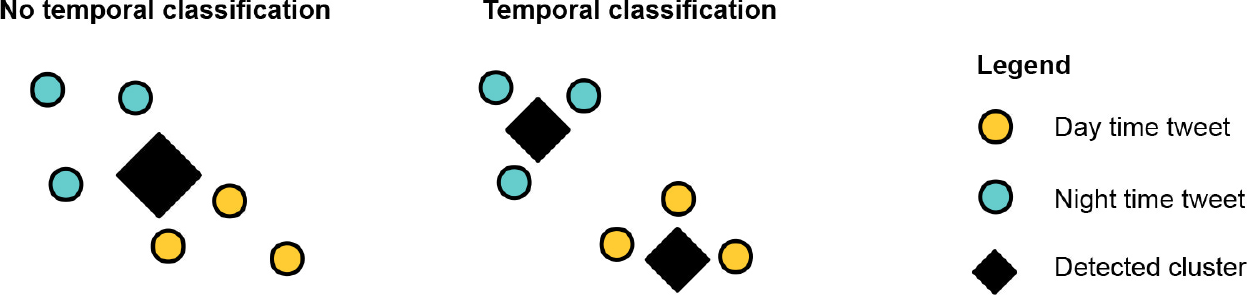}
\caption{Temporal classification of Tweets}
  \label{fig:temporal_classification}
\end{figure}

A possible approach would be to group the tweets in different subsets, as shown in Figure \ref{fig:temporal_classification}3. Six tweets are shown; three day-time tweets and three night-time tweets. Using no temporal information, a single cluster will be detected. Clustering subsets of tweets can lead to a higher number of users per cluster: tweets in each other vicinity might yield more than one cluster, as they would be in different subsets. It is challenging to determine whether clusters located in approximately the same location actually belong to the same activity location. 

\section{Results}\label{sec:results}

\subsection{Detecting activity locations}

Due to social network data's nature, recognition of user's locations differs fundamentally from another frequently used location-based data source used in transport research: GPS data collected through smart phones or dedicated GPS trackers. While the latter data sources provide location and speed information, making it possible to perform not only mode detection, but also to detect start and end times of trips activities, social network data only shows when a user posts geo-tagged information to the social network. Individual locations could be obtained from social media by means of location check-ins; this, however, leads to a limited subset of users and location types \citep{NoulasEtAl_PLOS_2012,HasanSUkkusuri_TransResC_2014}. The challenge with a data stream coming from Twitter is to determine whether a user is at an activity location or en-route.

\begin{figure}
\centering
\begin{subfigure}[b]{0.45\textwidth}
    \includegraphics[width=\textwidth]{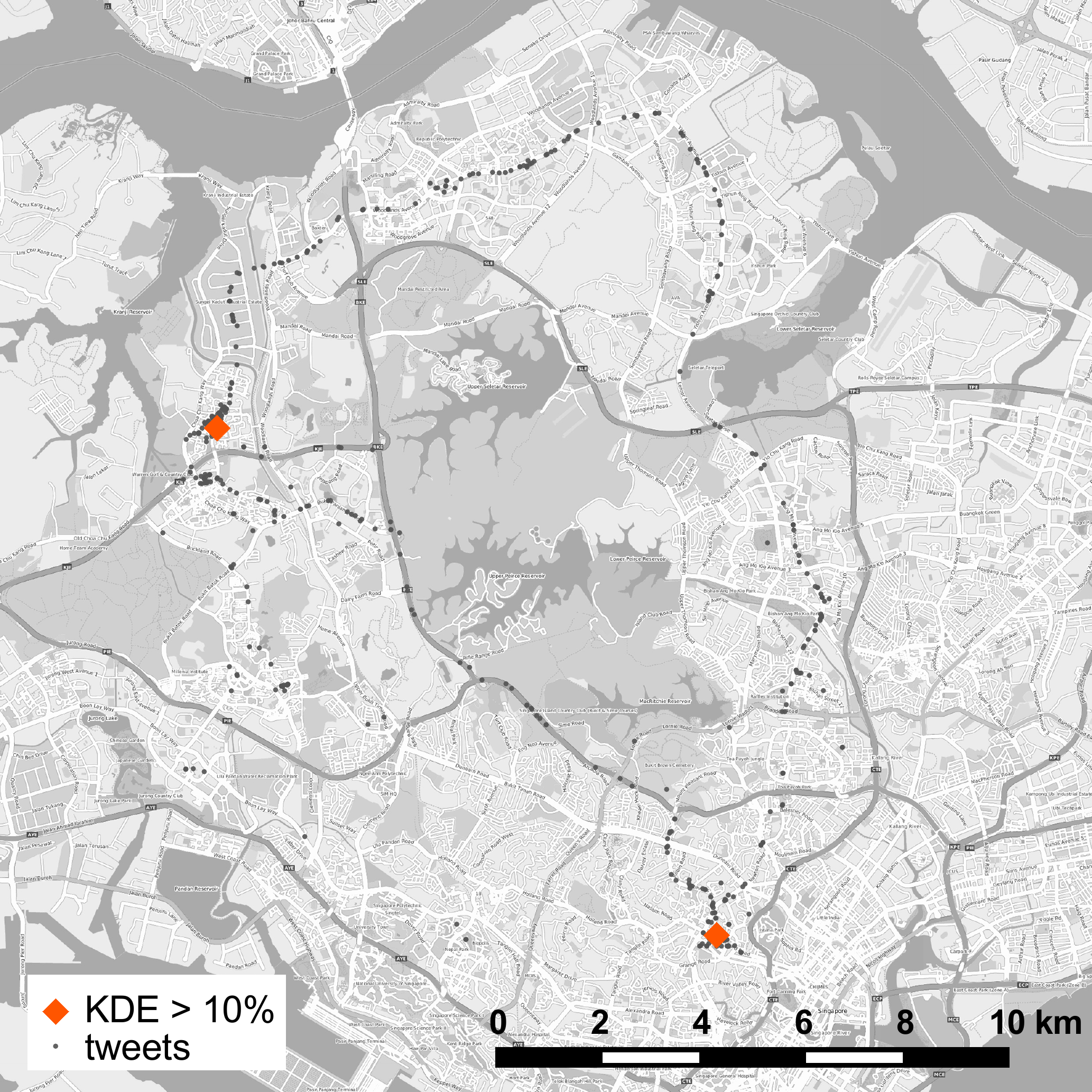}
    \caption{Kernel density estimation (KDE > 10\%)}
    \label{fig:kde}
  \end{subfigure}
  \begin{subfigure}[b]{0.45\textwidth}
    \includegraphics[width=\textwidth]{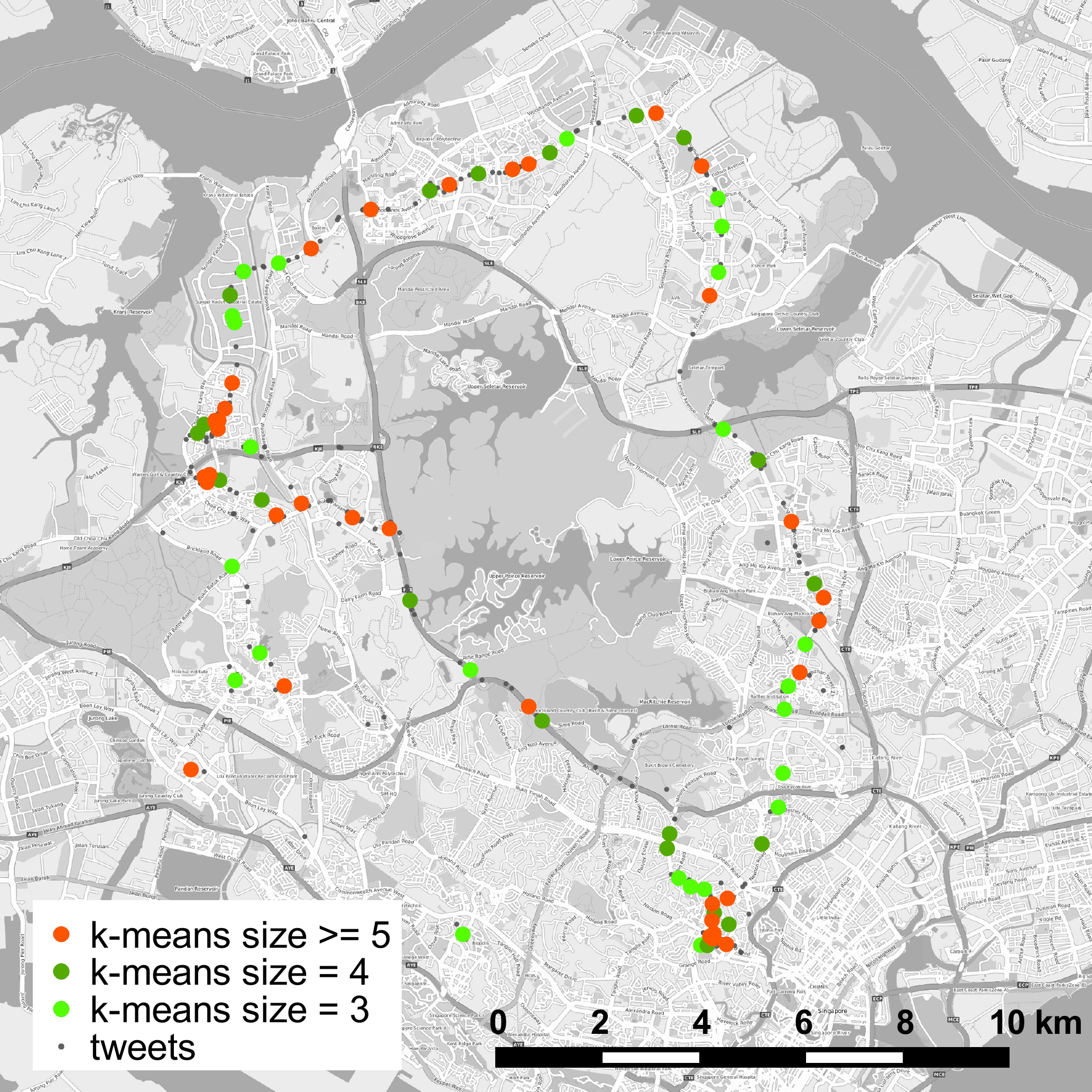}
    \caption{K-means cluster sizes 3, 4 and larger than 5}
    \label{fig:k3}
  \end{subfigure}
  \caption{This figure depicts 1,405 individual tweets of a randomly selected user and the detected clusters by means of k-means clustering and kernel density estimation. Background map: \protect\url{http://openstreetmap.org}. 
}
  \label{fig:random_user}
\end{figure}

Figure \ref{fig:random_user} shows the tweets of an active user. The selected user has tweeted 1,405 times over the course of the 8-month observation period. While the data might look similar to GPS data in terms of detected trajectories, these tweets are not necessarily ordered by time. The user's main locations have been identified by both k-means clustering and kernel density estimation (KDE). 

In k-means clustering, each data point (tweet) needs to be assigned to a cluster. A cluster recognized by k-means clustering has a certain number of tweets belonging to that cluster, which we will refer to as the cluster size. In the figure, clusters of different sizes are depicted, as well as clusters recognized with kernel density estimation. Using KDE, two clusters are recognized; with k-means clustering, more than 100 clusters are recognized containing two tweets or more. Figure \ref{fig:tweets_vs_kdc} in the appendix provides more detail on the number of tweets required to detect a certain number of clusters. 


To determine the merits of both the k-means clustering and KDE, both methods are evaluated by the number of clusters recognized per user and each cluster’s strength. Currently, the strength of each cluster is evaluated as follows: 

\begin {itemize}
  \item For clusters recognized by \emph{k-means clustering}, the strength is calculated as the number of tweets belonging to each cluster; the size of the cluster. A distinction is made between clusters having 1, 2, 3, 4 and 5 or more tweets. 

  \item For clusters recognized by \emph{kernel density estimation}, strength is calculated as the contribution (the level) of a single cluster to the sum of the levels of each cluster of a single user. Clusters contributing less than 5\%, 10\% and 20\% respectively to the sum of the levels are filtered out. 

\end{itemize} 

\begin{figure}
	\centering
	\includegraphics[width=\textwidth]{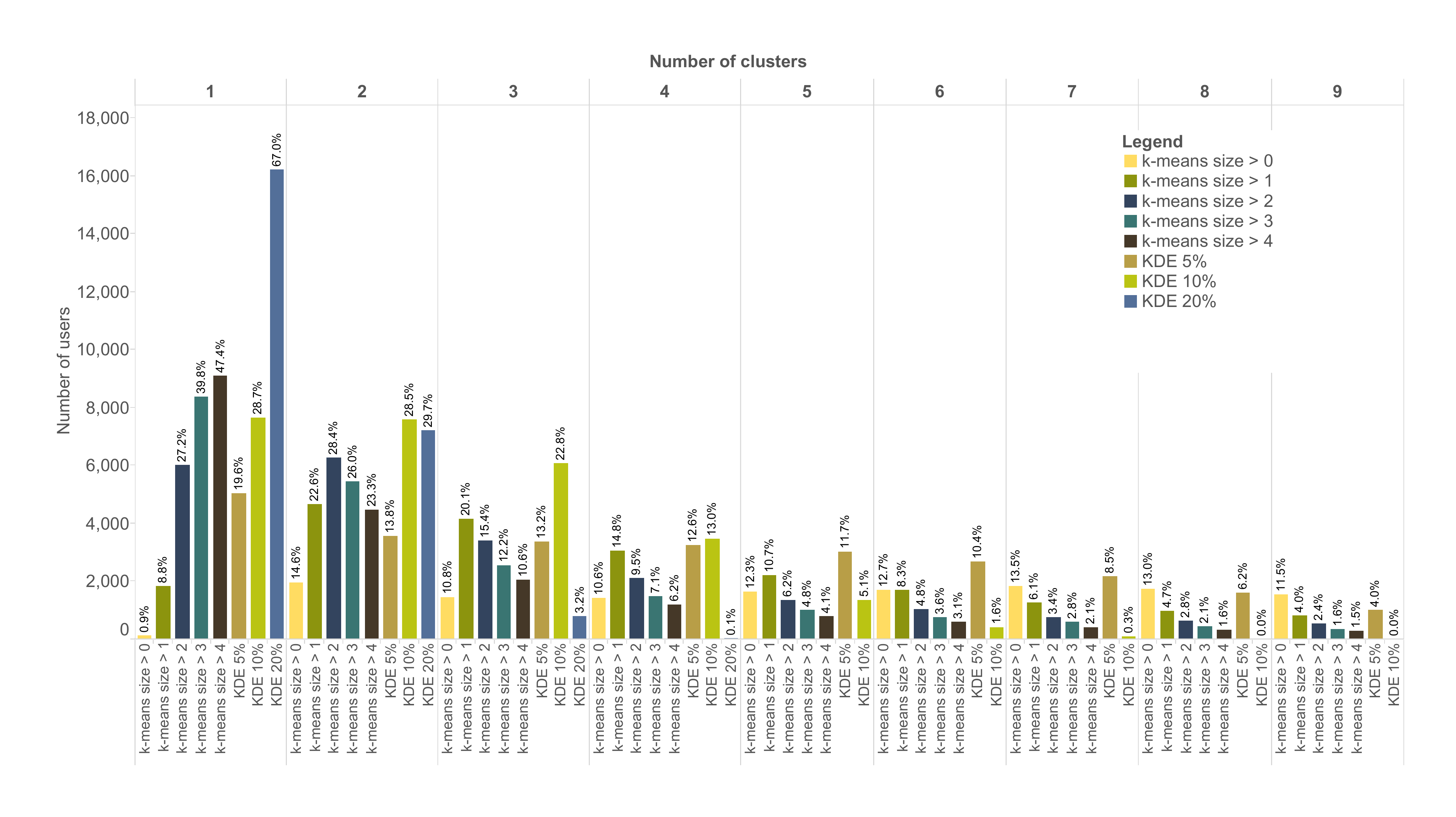}
	\caption{Number of clusters detected with different cluster methodologies}
	\label{fig:clusers_vs_kernel}
\end{figure}

The results of the evaluation are presented in Figure \ref{fig:clusers_vs_kernel}; results only include users tweeting in Singapore, or Singapore and overseas, tweeting 10 times or more. An intuitive result is found; if threshold levels for a cluster's strength are set low, the number of clusters found by both methods is high; when setting thresholds' value high, a lower number of clusters is detected. If a minimum cluster size of 4 is set for the k-means clustering, 47\% of the users have only 1 cluster and 23\% percent have 2 clusters. If a minimum contribution level of 20\% is set for KDE, 67\% of the users have only 1 cluster; while with a level of 5\%, 80\% of the users have more than one cluster.
From this, the relationship between the chosen threshold and the number of clusters becomes apparent. If the goal is to determine the number of frequently visited locations, thresholds will need to be set. However, if the goal is to determine a user's activity space, one can avoid setting thresholds, thus not deleting user information. 

K-means clustering can be considered the standard method for clustering. Visual inspection of our results per user and inspection of histograms indicate that k-means is not well suited to make a distinction between en-route events and activity locations. This follows from the sparse nature of the data and the requirement that each data point belongs to exactly one cluster. We will now proceed with clustering by means of kernel density estimation and a contribution level of 10\%. This means that at least 10\% of the tweets of a single user should be in a single cluster. This value was chosen, as it seemed to strike a balance between the filtering of en-route activities and stationary activity locations.


\subsection{Comparison with travel survey data}

\begin{figure}
	\centering
	\includegraphics[width=1\textwidth]{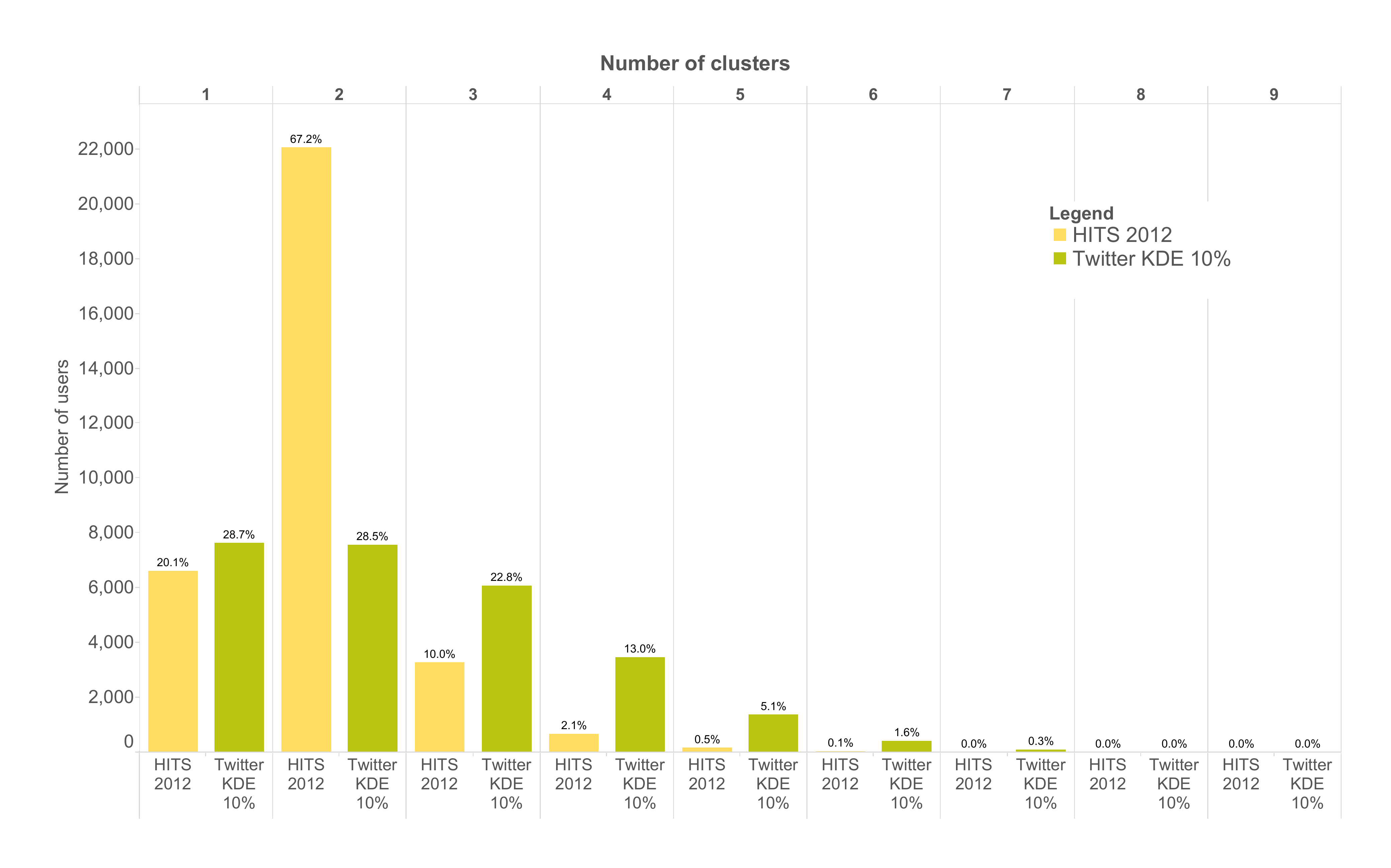}
	\caption{Number of reported activity locations in travel survey as compared to the number of clusters in Twitter by means of kernel density estimation}
	\label{fig:kde_vs_hits}
\end{figure}

\emph{Number of activity locations}
When comparing clusters detected from Twitter data with reported activity locations from the travel diary survey, we assume each reported trip indicates a separate activity location. Figure \ref{fig:kde_vs_hits}  allows for a comparison of one-day travel survey data and Twitter data collected over a period of several months. Survey respondents with only a single identified activity location (25\%) included retirees, homemakers and domestic workers. Over 50\% of the survey respondents report two activity locations. The applied cluster methodologies detect more activity locations than reported in the survey. For activity locations recognized with KDE and a contribution level of 10\%, 29\% of Twitter users have a single cluster, 28\% of the users have two, 23\% of the users have three clusters, 13\% of the users have four and 5\% of the users have five clusters. 

\begin{figure}
	\centering
	\includegraphics[width=\textwidth]{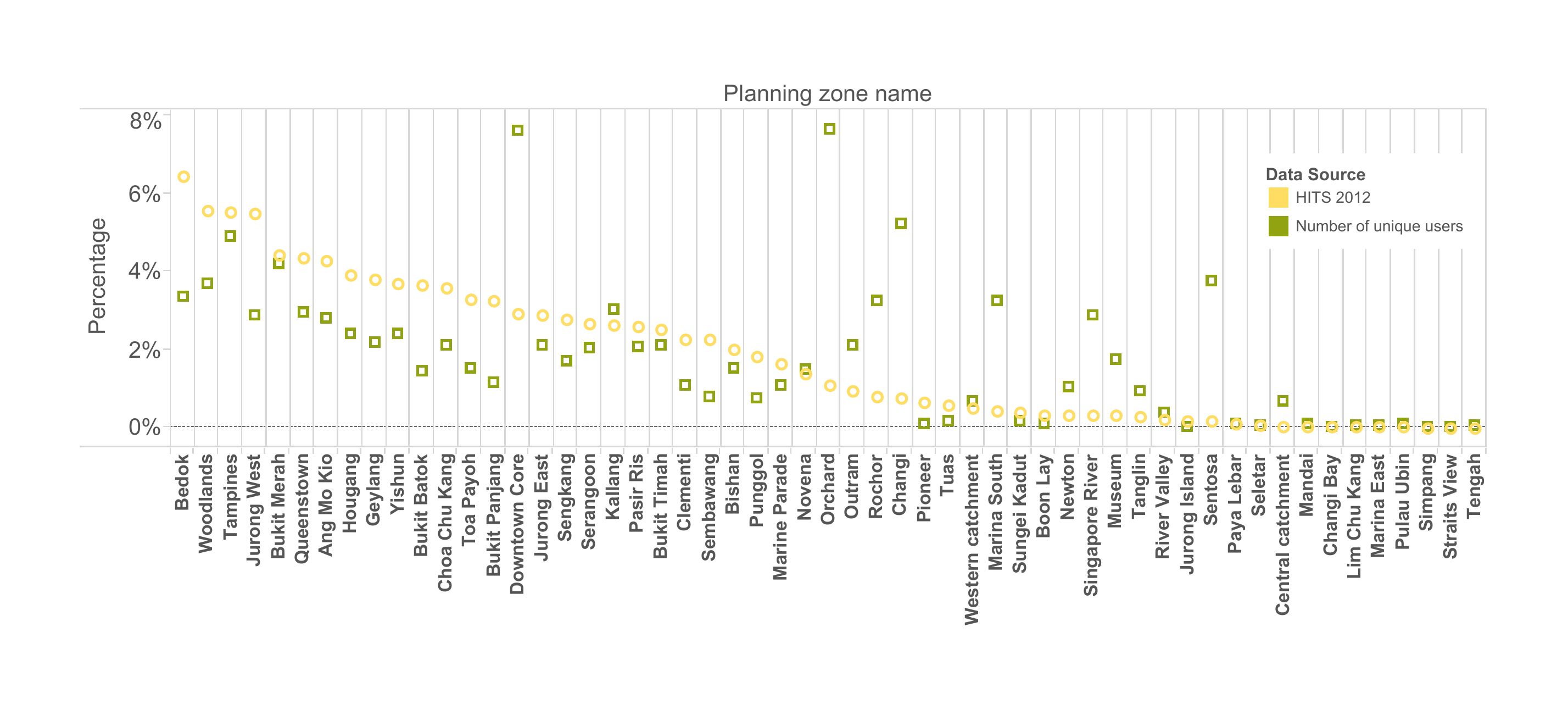}
	\caption{Comparison of clusters detected in Twitter and reported activity locations in the 2012 household interview travel survey}
	\label{fig:hits_vs_twitter}
\end{figure}

Figure \ref{fig:hits_vs_twitter} shows the number of Twitter clusters per planning zone versus the number of respondents in travel survey data with an activity location in the planning zone. A relatively high number of clusters was detected in the planning zones: Downtown Core, Orchard, Singapore River, Marina South, Museum and Sentosa. These zones typically contain numerous restaurants, as well as shopping opportunities and other leisure destinations. 


\emph{Distance comparison}

\begin{figure}
	\centering
	\includegraphics[width=\textwidth]{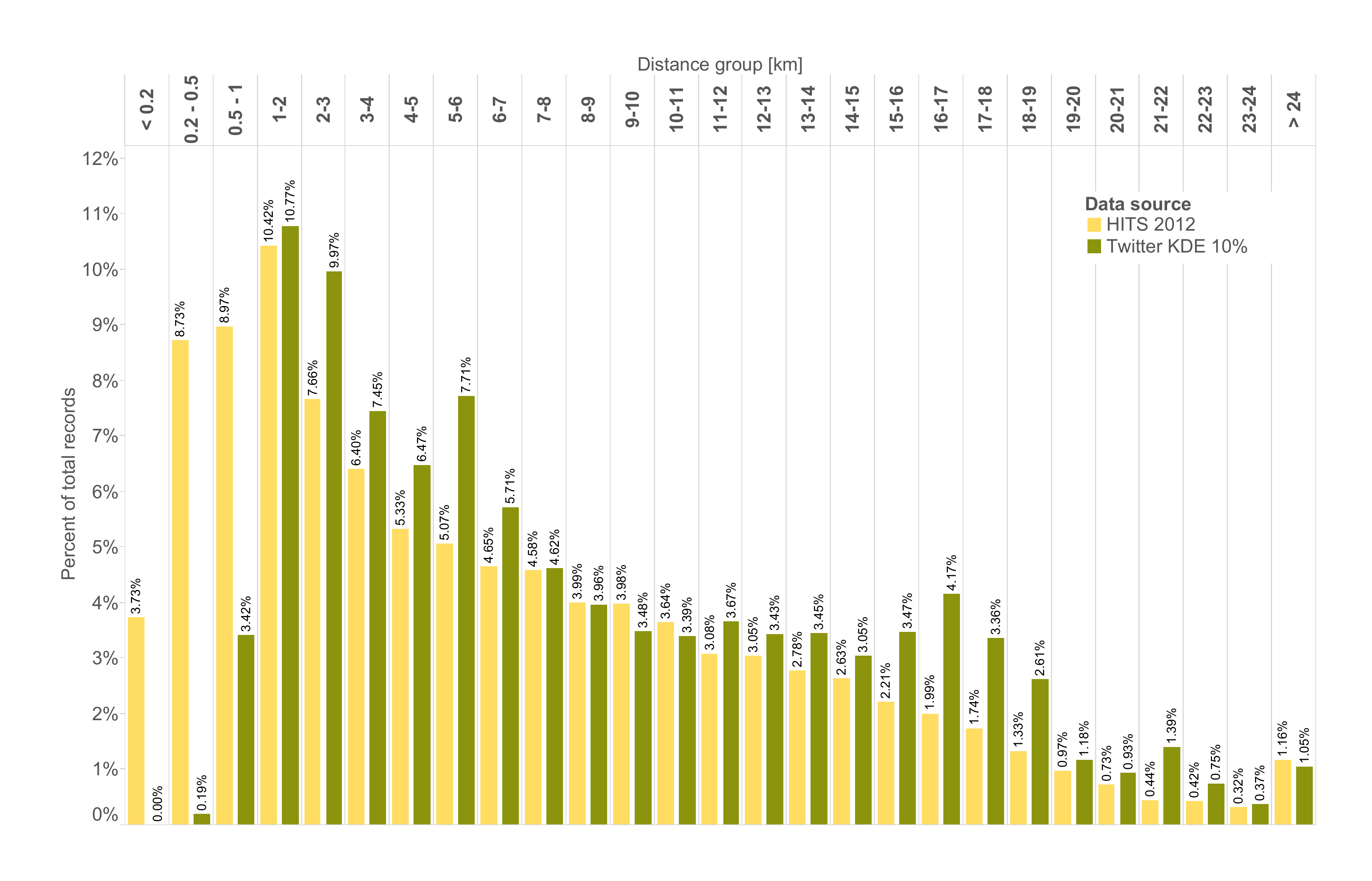}
	\caption{Comparison of distances between reported activity locations in the household interview travel survey 2012 and activity locations detected in Twitter using kernel density estimation (17,930 users).}
	\label{fig:distance_comparison_def}
\end{figure}

In addition to the visual inspection of clusters and assessing the total number of clusters per user, we compared distances between clusters detected in Twitter and distances between reported activity locations in the Singapore travel diary survey. To assess whether the distances between activity locations found in different data sources correspond for both data sources, Euclidean distances between all unique reciprocal locations per user were calculated. For example, if a user reported trips to three distinct locations (e.g. home, work, leisure), we calculated the distances between home-work, home-leisure and leisure-work. A similar procedure was followed for clusters detected in Twitter using KDE with a threshold value of 10\% as a reference case.
In Figure \ref{fig:distance_comparison_def}, distance comparison results are presented, showing that distances between detected clusters and reported activity locations match for most distance categories. In the household interview travel survey, a higher number of cluster-pairs was reported that were separated by less than 1 kilometer. A closer analysis of HITS reveals that clusters separated by less than 1 kilometer are often activity pairs 'home-education' (44\%), 'home-pick up drop' (11\%) and 'home-work' (10\%). Furthermore, clusters detected in Twitter data are more separated from each other than activity locations observed in travel survey data. This is especially notable for clusters separated by more than 15 kilometers. To determine whether the two distributions can be considered similar, we performed a Kolmogorov-Smirnov test. The test revealed that the two distributions were similar, with a p-value of 0.93. 
s

\subsection{Comparison with aggregated statistics}
Clusters detected through KDE and using a threshold of 10\% were compared to Singapore's population statistics \citep{SingStat_Webpage_2010} and estimated work locations  \citep{ChakirovErath_IATBR_2012,OrdonezErath_TRR_2013}. The results of the comparison are presented in Figure 9, showing the number of users with one or more clusters in each respective planning zone against the population (top) and the number of work locations (bottom). Bedok has a resident population of 589,038, according to the 2010 Singapore census; the number of detected work locations with smart card data in the downtown core is 185,000. Because the order of magnitude differs from the number of users found in Twitter, the population percentage residing and working in each zone is shown and compared to the percentage of unique Twitter users with a cluster in this zone. 

A direct comparison between activity locations detected from Twitter data and resident population and work activity densities is challenging. While KDE clustering seemed to reliably identify activity locations, it would require other analysis methods to deduce the type of activity performed at the identified location. On the whole, identified activity densities corresponded quite well with where people live or work. 

The percentage of detected clusters in various zones matched the population percentage in several planning zones, most notably in the planning zones Bukit Timah, Novena, Marine Parade, Kallang and Queenstown. The first three planning zones were characterized by a high percentage of private property and correspondingly higher income. A further distinction by age and income is necessary to further analyze Twitter clusters and socio-demographics. The Pearson correlation between the number of unique users and workplaces per zone is 0.29. The Downtown core has the highest number of work locations and the second-highest percentage of Twitter users. A similar observation can be made for other planning zones, such as Bukit Merah and Queenstown. The Pearson correlation between the number of unique users and workplaces per zone is 0.70. 

Several zones revealed a higher number of Twitter users compared to population and workplace figures. These are: the shopping district Orchard, the airport Changi, the leisure island Sentosa and the entertainment area Singapore River. This comparison highlighted that clusters where users tweet are not limited to work, home or discretionary activity locations, but occur in areas that are particularly popular for leisure activities. 

\begin{figure}
	\centering
	\includegraphics[width=\textwidth]{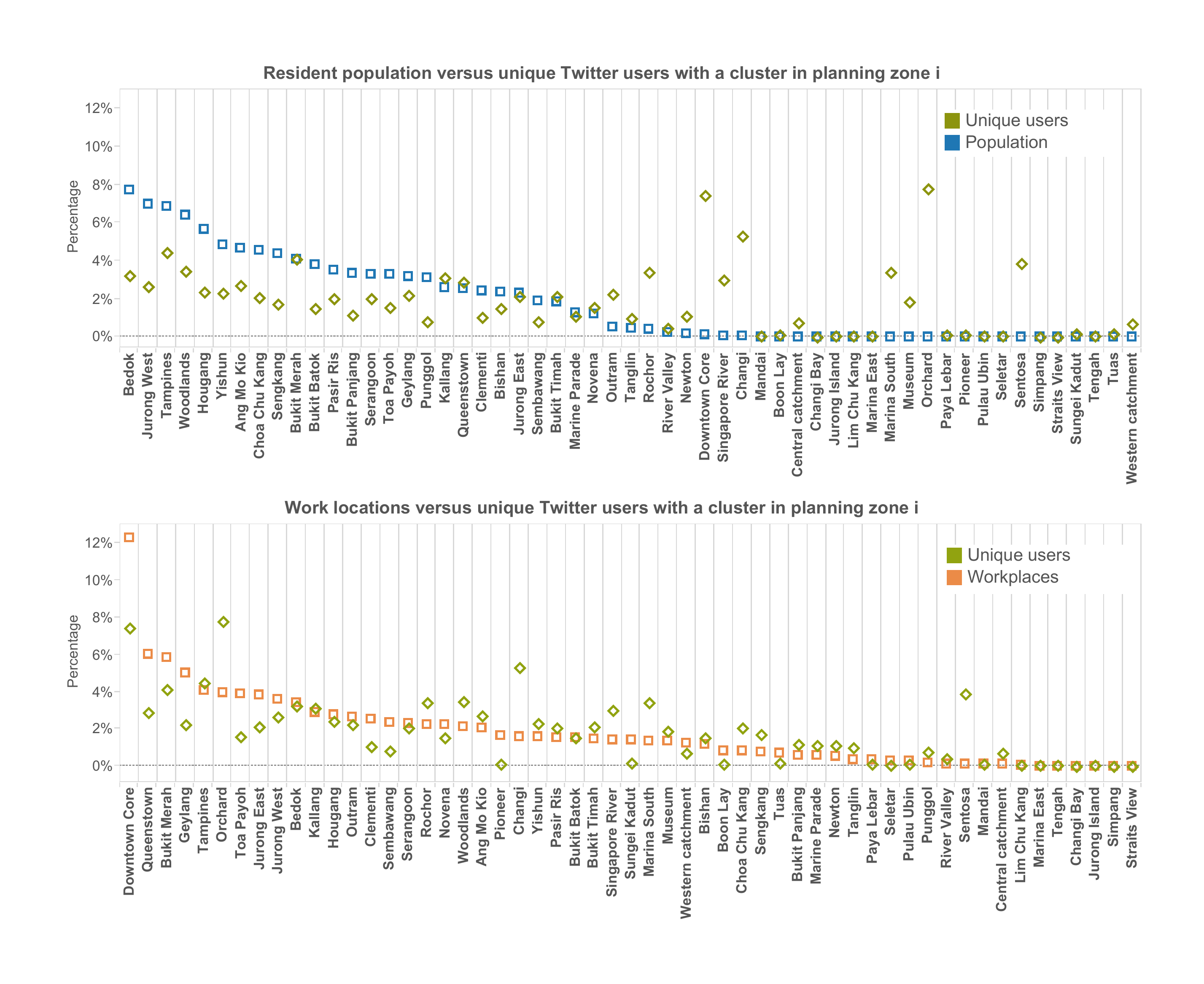}
	\caption{Percentage of unique users with a cluster in planning zone \emph{i} plotted against the population (top) and work locations (bottom). }
	\label{fig:twitter_vs_singstat}
\end{figure}

\subsection{Detecting transitions}
The third comparison involved comparing origin-destination (OD) matrices derived from public transport smart card data with transitions observed in Twitter data. OD matrices from smart card data were derived from journey start and end transit stops; no attention was paid to transfers. For instance, consider a user traveling from zone A to zone C with a transfer in zone B. The travel from A to C is considered a journey. However, the user could have transferred in zone B and is required to tap his card if the transfer involves a bus trip. As we have multiple days of smart card data, it is possible to observe multiple journeys from zone A to zone B. However, and to ensure comparability with Twitter data, each distinct zonal pair was counted once per user. A transition is defined as the relationship between zone A to zone C for a single public transport user.

To analyze Twitter data with similar definitions, we took - as a basis - clusters detected with kernel density estimation and apply a threshold of 10\%. Tweets located within the kernel contour were considered to be part of the cluster. Subsequently, each user's total tweets were ordered, by time, to determine common transitions between locations. A transition occurred when two subsequent tweets stemmed from different clusters. For instance, consider tweet $x$ occurring at $t_1$ in cluster 1, tweet $y$ occurring at $t_2$ in cluster 2 and tweet $z$ occurring at $t_3$ in cluster 1. The transition from cluster 1 to cluster 2 is counted as a transition. Similar to studies using call detail records (CDR), we were faced with the handling of the time between two  subsequent events, the inter-event time. In CDR data, a user's location is lost when the phone is not used and not all movement can be captured. The origins and destinations that can be captured as are referred to as transient origins and destinations \citet{WangPEtAl_SR_2012,IqbalEtAl_TransResC_2014}.

In order to analyze Twitter data according to similar definitions, we take as a basis clusters detected with kernel density estimation and apply a threshold of 10\%. Tweets located within the contour of the kernel are considered to be part of the cluster. Subsequently, all tweets of each user are ordered by time to determine common transitions between locations. A transition occurs when two subsequent tweets stem from different clusters. For instance, consider tweet $x$ occurring at $t_1$ in cluster 1, tweet $y$ occurring at $t_2$ in cluster 2 and tweet $z$ occurring at $t_3$ in cluster 1. The transition from cluster 1 to cluster 2 is counted as a transition. 
Similar to studies using call detail records (CDR), we are faced with the handling of the time between two subsequent tweets, the inter-event time. In CDR data, a user's location is lost when the phone is not used and not all movement can be captured. The origins and destinations, that can be captured as transient origins and destinations are referred to as transient origins and destinations \citet{WangPEtAl_SR_2012,IqbalEtAl_TransResC_2014}. 

Despite the fact that the exact OD pair might not be observed, it was still possible to capture a portion of it. In CDR data, different values for the median inter-event time were given, varying from 260 minutes to 500 minutes \cite{CalabreseEtAl_PC_2011}. The average inter-event time observed in the Twitter dataset was1576 minutes; the average user median was 1954 minutes. Given this long time span between two events, we chose not to use the inter-event time in this analysis. By discarding the time between two subsequent tweets in this analysis, transitions of users that tweet less frequently were included in this analysis. Also, locations where users tweet less frequently, such as the home location, can be included in the transition analysis. As such, all spatial fingerprints of users can be used, but the extracted OD-relations should be considered as trends and not exact figures. 

Per user, each distinct zonal pair is counted once. We thus assume that transitions will occur from time to time between these activity locations. One limitation of this approach was that other possible transitions of this user occurring outside the measured location-based social network (tweets) were not measured.

\begin{figure}
	\centering
	\includegraphics[width=\textwidth]{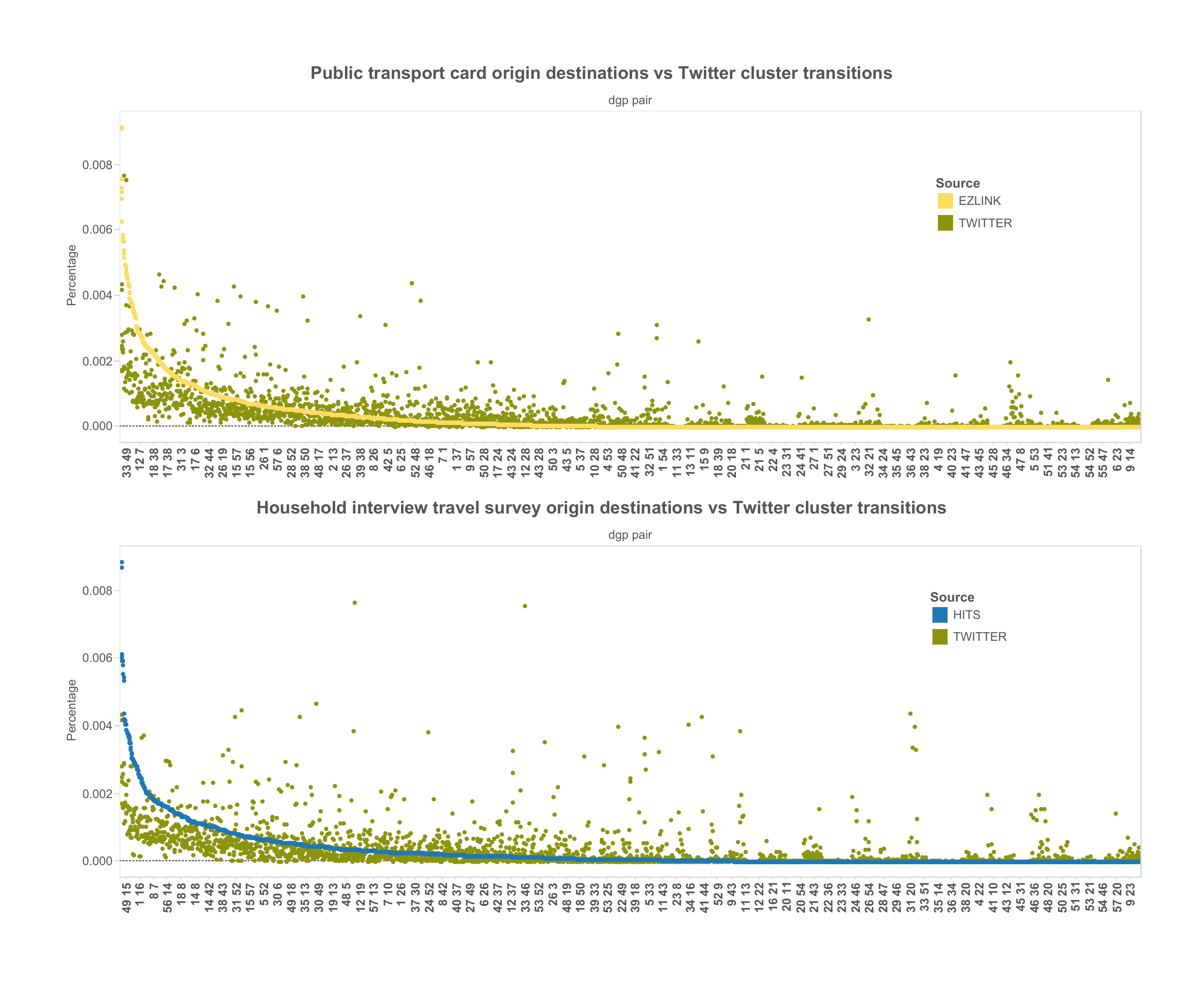}
	\caption{Transitions as calculated from Twitter versus weekday public transport smart card data (top) and household interview travel survey data (bottom) journeys per planning zone pair. Intra-zonal trips have been excluded. The relative flow per origin-destination pair is shown. Records are sorted by the percentage per OD-pair from public transport smart card data and household interview travel survey data, respectively. The numbers indicate that the zonal numbers.}
	\label{fig:transition_comparison}
\end{figure}
In Figure \ref{fig:transition_comparison}, transitions are calculated from detected locations with KDE and a threshold of 10\% versus public transport smart card data (top) and household interview travel survey data (HITS, bottom) per planning zone. Intra-zonal and weekend trips have been excluded; the latter were excluded because they do not occur often in travel survey data.

To compare results from both data sources, the relative flow per OD-pair is shown. Records were sorted by the percentage per OD-pair from smart card data and HITS data respectively. This approach made it possible to compare trends and detect differences between both data sources. In both cases, transitions derived from Twitter followed a trend similar to both smart card data and HITS. The correlation coefficient between HITS and smart card data was $0.859$ and the p-value associated was less than $10^{-3}$; the correlation coefficient between HITS and Twitter was $0.512$ and the p-value associated was less than $10^{-3}$ and the correlation coefficient between smart card data and Twitter is $0.655$ and the p-value associated is,  again, less than $10^{-3}$. While the correlation between HITS and smart card data is very high, the correlation between HITS and Twitter data is substantially lower, but statistically still very significant. The time between transitions in Twitter data can amount to multiple days. These transitions would not be recorded in a single day travel survey. 

\begin{table}
  \centering
  \caption{Breakdown of the number of users applying a kernel density estimation with a threshold of 10\%. The table shows if a user only has clusters in Singapore or both in Singapore and overseas.}
    \begin{tabular}{rrrr}
    \toprule
    \textbf{Country} & \textbf{Region} & \textbf{Only Singapore} & \textbf{Singapore and outside Singapore} \\
    \midrule
    Singapore & Singapore - all & 14,628 & 3,944 \\
    Malaysia & Johor Baharu &       & 1,517 \\
          & Other Malaysia &       & 39 \\
    Indonesia & Batam &       & 426 \\
          & Other Indonesia &       & 27 \\
    Thailand & Thailand - all &       & 67 \\
    \bottomrule
    \end{tabular}%
  \label{tab:breakdown_users}%
\end{table}%

An advantage of social network data is that the collection of data is not limited by geographical boundaries. Number of users and tweets in Singapore and outside of Singapore was presented earlier in (Table \ref{tab:general_overview}). In Table \ref{tab:breakdown_users}, a breakdown of user numbers is presented with clusters only in Singapore, and in Singapore, Malaysia, Indonesia and Thailand. Clusters are detected with KDE and a threshold of 10\%. Almost 4,000 users have a cluster in Singapore and outside of Singapore. The majority of these users have one or more clusters in the province adjacent to Singapore, Johor Bahru. Johor Bahru, a city, is across the border from Singapore in Malaysia and is accessible by foot, car, motorcycle and frequent bus services; Batam is the most populated island within the archipelago of the Indonesian province Riau and only accessible by ferry.
\section{Discussion \& Outlook}\label{sec:discussion}
This paper highlighted the possibilities of using social network for transport demand models. Factors including detection of an individual's activity locations, spatial separation between these locations and transitions between these clusters were investigated. Whereas previous work \citep{HasanSEtAl_UrbComp_2013,NoulasEtAl_PLOS_2012,HasanSUkkusuri_TransResC_2014}considered only a subset of this data -location check-ins - we included all available data. Twitter is sometimes considered ’Big Data’. This can be considered relative to other data sources such as the number of records obtained from dedicated GPS trackers. For Singapore, we counted approximately 45,000 unique users who tweet at least 10 times or more in a time span of 8 months and correspond to around 0.8\% of Singapore’s resident population. These users tweet 2 million times in total. This number of users is comparable to the usual sample size of a travel diary survey, but the observed time span as compared to multi-day and multi-week survey and the number of activity location data points is several orders of magnitudes larger and hence can be sensibly characterized as 'big'. 

To determine an individual’s locations, two clustering methodologies were applied. A first challenge lay in the distinction between en-route Twitter events and Twitter events at activity  locations. The application of kernel density estimation for the detection of clusters, as proposed by \cite{LichmanSmyth_KDD_2014}, yielded more promising results than k-means clustering, which still can be considered the standard method for clustering. The kernel density approach required a bandwidth $h$. Setting a high value for $h$ can result in over-smoothing. Translated to the detection of activity locations, this can result in a lower amount of detected locations in each others proximity. In this study we applied a bandwidth of 200 meters, roughly corresponding to an area of 500 meters around the KDE result, considered a reasonable setting considering positioning errors in a dense urban environment. In the case of GPS data, speed information can be used to detect activity locations. In the case of Twitter data, no speed information is available.

Therefore, we determined the strength of a detected cluster relative to a user's other clusters and apply a threshold. This threshold determined whether a detected cluster was an activity location or en-route event. By applying this threshold, less frequently visited locations were filtered out. Despite this filtering, more locations were detected with all threshold settings than reported in survey data. 

An important input for transport demand models was the distribution of trip distances. Spatial separation between detected clusters from Twitter and reported activity location from travel survey diary data corresponded well. Short trips under 1 kilometer, - 44\% of which were home-school 
 trips – were under-estimated. Whether this was due to over-smoothing, or the fact that primary school students are not active on Twitter, is open for discussion. 



Theoretically, detected activity locations could be used for trip distribution purposes (e.g. gravity model estimation). However, by just considering a user's Twitter locations, no information on trip purpose was known. By combining the temporal component of Tweets, rules incorporating time and text recognition the type of activity could be inferred. Also, redefining the threshold definition for KDE clustering gave an indication of the frequency with which a location was visited, an interesting topic for further research. More difficult to infer would be an individual's mode and route choice. For very frequent Twitters users, several paths can be observed between detected clusters. However, to infer routes from these tracks would require a wide range of assumptions and rules. 

An important input for activity-based models was the range of activity locations an individual frequents. While it seemed promising to use the detected locations for activity chain generation, inferring activity start time, activity end time, activity duration and sequence can be challenging due to the scattered nature of social network data. However, the number of locations detected besides main clusters such as 'home' and 'work' gave insight in spatial dispersion (i.e. distances) of these additional clusters and opens up possibilities for multi-day activity based  models and the sampling of discretionary activity locations. \cite{SchoenfelderAxhausen_KollSchretzenmayrEtAl_2004} discussed several methods to calculate activity spaces. Despite the varying nature of different data sets, comparison of different activity spaces offers another method to compare different data sources. 

On an aggregated level, two comparisons were made. From the first comparison between the detected cluster and population statistics it was deduced that Twitter events occur less frequently at home locations and/or that Twitter users form only a sub-sample of the population; in several homogeneous planning zones, a good match between detected clusters and population statistics was observed. A further distinction of population statistics by age and income remains for further work. A similar trend could be observed when comparing detected locations with work locations. The correlation between work locations and detected clusters was higher than the correlation with population statistics. Several zones, the main shopping area Orchard, the airport Changi, the entertainment area Singapore River and leisure island Sentosa showed more detected locations. Another topic for further work is the inclusion of the temporal component in the clustering algorithm \cite[e.g.][]{HasanSUkkusuri_TransResC_2014}. By combining the temporal component and a wider range of zonal statistics, it would be possible to estimate what contributes to Twitter usage per zone. 

The second comparison considered transitions between zones as detected in Twitter data and transitions as observed in public transport smart card data and survey data. The high correlation between the Twitter data and the two traditional data sets opens promising avenues for future research. The time between two subsequent tweets was not included in this analysis, as the time between two tweets - the inter-event time – was high when compared to other data sources, such as call detail record data. The latter data sources included more data points per user, as it contained voice and data transactions, not just a subset of transactions, as is the case with Twitter data. However, transitions computed with Twitter can be considered a proxy between a user's main activity locations. 

Because Twitter data is available throughout the world, it presents an opportunity to compare how distances between activities are distributed for different cities and to relate the findings to the respective spatial extents and population densities. Another promising application could be the detection of long-distance and cross-border trips, which are difficult to observe representatively in travel diary surveys and are seldom covered in public transport smart card records. The availability of airline ticket records would allow similar comparisons to those presented in this paper, to test how well Twitter data represents actual long distance travel behavior.

Despite these unanswered questions, location-based social network data provides a promising data source for the detection of activity locations and analysis of mobility patterns, especially considering the potential to track users over a longer span of time against negligible costs. Social network data can give a first impression about the prevalent distribution of activity locations and trips distances. This is particularly useful for regions which lack mobility diary surveys, but feature an active Twitter user base. However, Twitter data does not document socio-demographic user characteristics or context of the observed activity locations. Although Twitter data allows identification of a user's main activity locations, it can not replace traditional travel surveys or smart card data and provide a similar level of understanding of travel behavior. However, given the limitations of those approaches to represent cross-border flows and long distance trips, Twitter data can still be considered a valuable supplement to more traditional travel surveys, especially as the accessible user base is many times larger. We also see potential to better understand travel behavior in new mobile phone applications such as Strava, Moves, or human.co, which are designed to track individual movement patterns.

\bibliography{all-eng}

\section*{Acknowledgements}
The research conducted at the Future Cities Laboratory is co-funded by the Singaporean National Research Fund and the ETH Zurich, and located at the Campus for Research Excellence And Technological Enterprise (CREATE). Data61 is funded by the Australian Government. The authors would like to thank Professor Kay W. Axhausen for his suggestions and comments.  We wish to express our gratitude to the Land Transport Authority for providing us invaluable data sets on transport in Singapore. Also we are very thankful to the Urban Redevelopment Authority for providing us with a wide range of data sets.

\newpage

\section*{Appendix}\label{sec:appendix}
\begin{figure}
	\centering
	\includegraphics[width=\textwidth]{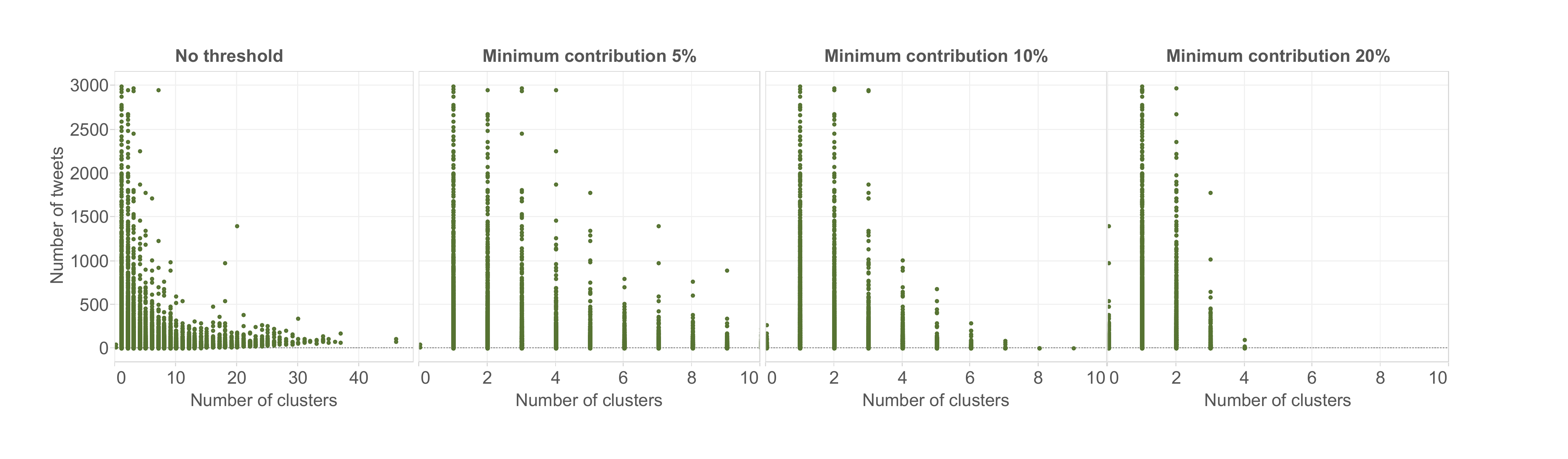}
	\caption{Depicted is the number of tweets versus the number of clusters detected; each point represents a user. The left-most scatter plot shows the case where no threshold is set for the contribution of a single cluster to the total level of user; the three other scatter plots present the results for a 5\%, 10\% and 20\% threshold respectively.}
	\label{fig:tweets_vs_kdc}
\end{figure}

One of the advantages of social network data is that the costs of collecting records for a longer time span are virtually free. Figure \ref{fig:tweets_vs_kdc} shows the number of tweets collected, as a proxy for time, versus the number of clusters recognized with different thresholds for the number of clusters when using KDE. Only users tweeting 10 times or more have been included. The left-most plot shows a counter-intuitive result: despite the high number of tweets not a high number of clusters is recognized. The three other plots show the number of clusters detected with different thresholds for the level of contribution. While a high number of tweets is required to detect 1 or more clusters, the effect of a high number of tweets per user on detecting the number of clusters per user is limited.

\end{document}